Temporal variation of Earth-based gravitational constant measurements

Jean Paul Mbelek

Abstract : As one knows, strong discrepancies are found between the different precise measurements of the gravitational constant carried out in Earth-based laboratories. While the precision are increasing in different laboratories and with various methods, these measurements are even more and more discordant. We have shown since 2002 that an improved 5D Kaluza-Klein (KK) theory may provide a satisfactory explanation to these discrepancies by referring to the geomagnetic field as a possible cause. Here we take advantage of different precise measurements performed at the same location but at different epoch to address the temporal variation of the gravitational constant measurements.

I-Introduction

The gravitation constant, G, is defined from Newton's universal law of gravitation which states that the force, F, that translates the gravitational pull between two bodies of mass m and M separated by a distance d is given by $F = GMm/d^2$. Nowadays, gravity is viewed as a gauge theory based on the group of diffeomorphisms, namely general relativity (GR). Thus, the gravitational constant appears in GR through the Einstein field equations. Furthermore, according to the renormalization group the coupling constants of renormalizable gauge theories should be looked at as running coupling constants and as such depend on the energy scale and the momentum transfer solely. However, gravity is recognized as a non-renormalizable theory, thus the question arises as to whether G is really a constant. It is usually believed that it should be so far below the Planck energy. In any case, the coupling constants of gauge theories do not depend on spacetime coordinates. However, multidimensional theories of gravitation which aim to unify GR and the gauge theories of the standard model of particle physics imply effective coupling constants after dimensional reduction[1]. This means that the genuine coupling constants, as defined in the bulk, cannot be directly measured in our 4D brane but derive through relations involving a 4D internal scalar field, $\phi$. Thus, the 5D KK theory implies an effective gravitational constant $G_{eff} = G/\phi$ and an effective fine structure constant $\alpha_{eff} = \alpha/\phi^3$. Now, the genuine 5D KK theory does not yet allow any quantitative difference on the experimental ground between these effective constants and their genuine counterparts. An improved 5D KK theory allows in contrast to make a noticeable difference between these effective and genuine constants. In our seminal paper [1], we addressed the issue of the discrepant G measurements. However, we did not deal with the temporal variation of the geomagnetic field. The main reason for doing so is that the gravitational constant measurements are usually averaged on time. However, we did mention that for those measurements that would not be averaged on time, one should expect small temporal variations of the value, $G_{lab}$, of the gravitational constant derived from Earth-based laboratory measurements. We claimed that the irreducible part of these variations might be related to both the Sq and the L field disturbances of the geomagnetic field. Actually, periodic variations of $G_{lab}$ with the lunar or diurnal period have yet been pointed out in the literature (see [2, 3, 4, 5]). It is presently believed that they are related to tides, but it could be made more effective through the temporal variation mentioned above. We notice that the gravitational measurements of ref. [6] are consistent with an annual variation. Recently, Anderson et al. have suggested a correlation between $G_{lab}$ and the length of day (LOD) [7] without invoking any physical mechanism but some unidentified systematic errors on the measurement process. Their fitting model is of the sinusoidal variation $G_{lab} = G + a \cos[(2\pi t/T) + \varphi]$ or a two-period fit $G_{lab} = G + a_1 \cos[(2\pi t/T_1) + \varphi_1] + a_2 \cos[(2\pi t/T_2) + \varphi_2]$ for a better fit ; $T = (5.899 \pm 0.062)$ yr, $G = (6.673899 \pm 0.000069) \times 10^{-11}$ SI, $a = (1.619 \pm 0.103) \times 10^{-14}$, $\varphi = 80.9°$, $T_1 = (5.911615 \pm 0.000028)$ yr and $T_2 = (1.023087 \pm 0.000042)$ yr. Let us notice that the authors finally pointed out that "there might be correlations with terrestrial magnetic field measurements". Besides, after enlarging the $G_{lab}$ data base and a few corrections to the time interval the measurements were carried out, Schlamminger et al. concluded that this significantly weakens the correlation to the LOD [8]. Besides, based on the aforementioned hypothesis, L. Iorio could predict for the LAGEOS satellite an orbital increase as large as 3.9 m yr$^{-1}$ in contrast with the observed decay of $-0.203 \pm 0.035$ m yr$^{-1}$ and an anomalous perihelion precession as large as 14 arcseconds per century for Saturn [9]. Moreover, these claims have been disputed by other authors [10, 11, 12]. Nevertheless, four most precise laboratory measurements of the gravitational constant have been performed at the same location[2], HUST (Huazhong University of Science and Technology in Wuhan, China), but at different epoch since 1997. Although these gravitational constant measurements are averaged on time too, they have yielded discordant values. It is tempting to suggest that these might be rather strong arguments in favor of a secular variation. Still nowadays, although numerous gravitational constant measurements have reached a relative uncertainty of about 10 ppm (13.7 ppm, 11.64 ppm and 11.61 ppm respectively at Washington [13] and HUST 2018 with respectively the TOS and AAF methods [14]) to less than 150 ppm in many laboratories, most of them differ widely from each other up to 550 ppm. Unless all these discrepant gravitational constant measurements just reflect mundane sources of error, the 5D KK theory stabilized by an external bulk scalar field (KKψ) that we proposed in 2002 seems to be a fine causal solution to the latter puzzle which otherwise seems to deepen from year to year. Let us recall briefly the motivation for the KKψ theory and its main features. As one knows the Lagrangian density of the 5D KK theory reads in the Jordan-Fierz frame $L = (-g)^{1/2} [(c^4/16\pi G) \phi R - ¼ \phi^3 \varepsilon_0 F^{\mu\nu} F_{\mu\nu}]$, which shows that the 5D KK theory is equivalent to a $\omega = 0$ generalized Brans-Dicke (BD) theory [15]. Now, P. D. Noerdlinger has shown, based on an argument first put forward by L.D. Landau and E. M. Lifshitz [16], that the stability of the Lagrangian density of the BD theory requires $\omega > 0$ [17] thereby proving the instability of the genuine 5D KK theory. In this respect, the KKψ relies upon an external bulk scalar field, ψ, in order to stabilize the 5D KK theory. The source term of the ψ-field, J, includes the contributions of the ordinary matter, of the electromagnetic field and of the internal scalar field, $\phi$. For each, the coupling is defined by a function of both scalar fields and it is also temperature and ambient matter density dependent, namely $f_X = f_X(\psi, \phi)$, where the subscript X stands for "matter", "EM" and "$\phi$". In order to recover the Einstein-Maxwell equations in the weak fields limit, these three functions are subject to the conditions: $f_{EM}(v,1) = f_{matter}(v,1) = f_\phi(v,1) = 0$, where v denotes the vacuum expectation value (VEV) of the $\psi$-field. The contributions of matter and $\phi$ are proportional to the traces of their respective energy-momentum tensors. Since the energy-momentum tensor of the electromagnetic field is traceless, a contribution of the form $\varepsilon_0 f_{EM} F_{\alpha\beta} F^{\alpha\beta}$ accounts for the coupling with it. The fit of our model to the data shows that $(\partial f_{EM}/\partial \phi)(v,1) v \gg 4\pi G/c^4$, as it can be expected near the vacuum at low temperature or high matter density [1,18]. However, we may suspect that $(\partial f_{EM}/\partial \phi)(v,1) v \leq 4\pi G/c^4$ at high temperature or low matter density. In short, the $\psi$-field couples more strongly to the other matter-energy sources in region of condensed ambient matter than in region where it is not. Apart from the gravitational constant measurements, the KK$\psi$ theory predicts and explains anomalous torque observed in the laboratory [19], provides a possible explanation to the anomalous thrusts observed in asymmetric resonant cavities [20, 21] and has been successfully applied to some astrophysical [22] and cosmological contexts [23]. However, some authors argued that scalar-tensor theories generically violate the weak equivalence principle (WEP). Thus, twelve years ago, the KK$\psi$ theory was harshly criticized by A. Rathke [24] who argued that the KK$\psi$ theory is nonviable. This author claimed that the computations he had performed in the framework of the KK$\psi$ theory lead to a violation of the WEP by four orders of magnitude for torsion-balance experiments. At that time, we gave an answer that questioned the physical basis for such computations [25]. Recently, taking a fresh look at reference [24], we have found that the KK$\psi$ constant used by A. Rathke for his computations (see Eq. 27 of ref. [24]) is actually six orders of magnitude greater than the one which is really obtained from the fits (see Eq. 19 of ref. [1]) to $G_{lab}$ versus the magnetic potential, V. Indeed, A. Rathke picked up the right value $\mathcal{F}^{-1} = (\partial f_{EM}/\partial \phi)(v, 1) v = (5.44 \pm 0.66) \times 10^{-6}$ fm/TeV obtained from the fits of G, but by converting the latter into SI units, he made a huge mistake by six orders of magnitude resulting in $\mathcal{F}^{-1} = (3.40 \pm 0.41) \times 10^{-8}$ m/J instead of $\mathcal{F}^{-1} = (3.40 \pm 0.41) \times 10^{-14}$ m/J. Now, all the computations of A. Rathke were based on SI units. Again, we have checked his computations using the wrong value $\mathcal{F}^{-1} = (3.40 \pm 0.41) \times 10^{-8}$ m/J in SI units and found the same values as displayed in table 1 and Eq. 48 of ref. [24]. Finally, A. Rathke intended to prove that the KK$\psi$ theory does not pass the test of the WEP. However, it turns out that even relying on the physical basis as he had suggested, the KK$\psi$ theory yields values that are two orders of magnitude below the current experimental limit of the WEP from Earth-based laboratory [26]. The last test of the WEP performed in space by the MICROSCOPE satellite has confirmed the WEP with a relative uncertainty of $2 \times 10^{-14}$ [27]. However, the coupling functions[3] $f_X$ and their derivatives cancel out in free-space because of its low matter density, in this way the space based laboratories (e.g., the MICROSCOPE, LAGEOS or SPOT satellites) cannot really constraint the KK$\psi$ theory. Another criticism has been put foreward by F. O. Minotti concerning the huge force that could be implied by the magnetic field of the Earth. Actually, this flaw is easily removed in linear homogeneous isotropic media by setting[4] $\beta$ ($\partial f_{EM}/\partial \psi$) (v,1) = − ($\partial f_{EM}/\partial \phi$) (v,1), where $\beta$ = $\beta(1/\epsilon\mu c^2)$ is the constant that appears in the equation of motion of a classical neutral test body ($u^\nu \nabla_\nu$) $u_\mu$ = ½ $\beta$ ($\partial_\mu \psi$ − $u_\mu$ d$\psi$/ds) and that in addition we subject to the constraints $\beta(1)$ = $\beta(0)$ = 1 and $\beta'(1)$ = $\beta'(0)$ = 0 for the first order derivative (see [28], for further discussion). Let us emphasize that the latter constraints are quite consistent with the anomalous torque and thrusts observed in the laboratory since the Maxwell invariant reads $\epsilon_0$ $F_{\mu\nu}$ $F^{\mu\nu}$ = 2 (**B.H** − **E.D**) in a dielectric or magnetic medium instead of $\epsilon_0$ $F_{\mu\nu}$ $F^{\mu\nu}$ = 2 ($B^2 c^2 - E^2$) as in the vacuum.

---

[1] Let us emphasize that the four spacetime dimensions of GR may be not compactified because the group of diffeomorphism defines a non-compactified gauge group unlike the Lie groups U(1), SU(2) and SU(3) which are compactified gauge groups and as such are associated to compactified extradimensions.

---

[2] As yet, the HUST team has published five values but the second one, HUST 05, is just a correction to the first, namely HUST 99.

---

[3] The coupling functions, $f_X$, depend, in the same manner as the external scalar field potential, on the ambient temperature, T, and the internal chemical potential, $\mu$, which increases with the density of particles. These coupling functions decrease with respect to T but increase with respect to $\mu$ on account that the latter quantity decreases with temperature.

II-A solution to the temporal variation of $G_{lab}$

In our previous work [1,18], we compared two hypotheses, namely the null hypothesis which assumes that one is indeed measuring the true gravitational constant, G, and the non null hypothesis which assumes that one is actually measuring an effective gravitational "constant", $G_{eff}$. Hereafter, we shall consider the temporal variation of the geomagnetic field. The fields' equations to solve for the geomagnetic potential V and the $\phi$-field are the following

**B** = − $\nabla$V(**r**,t) and $\Delta$V(**r**,t) = div **B** = 0, (1)

□$\phi$(**r**,t) = 2$\mathcal{F}^{-1}$ **B**$^2$/$\mu_0$. (2)

In the first order approximation, the solutions of Eq.(1) and Eq.(2) read respectively (see the appendix)

V(**r**,t) = ($a^3/r^2$) [$g^0_1$(t) cos$\theta$ + $g^1_1$(t) sin$\theta$ cos$\varphi$ + $h^1_1$(t) sin$\theta$ sin$\varphi$] ≈ V(**r**,$t_0$) + V̇(**r**,$t_0$) $\Delta$t, (3)

$\phi$(**r**,t) = 1 − $\mathcal{F}^{-1}$ V(**r**,t)$^2$/$\mu_0$, (4)

since $\Delta$V̇(**r**,$t_0$) = $\Delta$V(**r**,$t_0$) = $\Delta$V(**r**,t) = 0 and lim$_{r\to\infty}$ $\phi$(**r**,t) = 1,

V(**r**,$t_0$) = ($a^3/r^2$) [$g^0_1$($t_0$) cos$\theta$ + $g^1_1$($t_0$) sin$\theta$ cos$\varphi$ + $h^1_1$($t_0$) sin$\theta$ sin$\varphi$] (5)

and we have set

V̇(**r**,$t_0$) = ($\partial$V/$\partial$t)(**r**,$t_0$) = ($a^3/r^2$) [$\dot{g}^0_1$($t_0$) cos$\theta$ + $\dot{g}^1_1$($t_0$) sin$\theta$ cos$\varphi$ + $\dot{h}^1_1$($t_0$) sin$\theta$ sin$\varphi$], (6)

where, $\mu_0$ = 4$\pi$ × 10$^{-7}$ m kg s$^{-2}$ A$^{-2}$, a = adopted Earth radius = 6371.2 km (average distance from center to surface), r = a + h = radial distance from the Earth's center, h = altitude with respect to the geoid, $\theta$ = colatitude = 90° − Latitude, $\varphi$ = azimuth related to the longitude and $g^0_1$, $g^1_1$ et $h^1_1$ denote the Gauss coefficients (see ref. [29], for their IGRF or DGRF values) and t = $t_0$ + $\Delta$t.

---
[4] Thereby providing a value to ($\partial f_{EM}/\partial \psi$) (v,1) v which has hitherto remained undetermined.

Thus, by combining the effective gravitational constant and Eq.(4), one derives in the first order approximation

$$1/G_{eff}(\mathbf{r},t) = \phi(\mathbf{r},t)/G \approx 1/G - (1/G\mu_0\mathcal{F}) \, V(\mathbf{r},t_0)^2 - (1/G\mu_0\mathcal{F}) \, [2 \, V(\mathbf{r},t_0) \, \dot{V}(\mathbf{r},t_0) \, \Delta t + \dot{V}(\mathbf{r},t_0)^2 \, \Delta t^2] \quad (7)$$

$$1/G_{eff}(\mathbf{r},t) - 1/G_{eff}(\mathbf{r},t_0) \approx - (2/G\mu_0\mathcal{F}) \, V(\mathbf{r},t_0) \, \dot{V}(\mathbf{r},t_0) \, \Delta t, \quad (8)$$

in as much as $\Delta t \ll |V(\mathbf{r},t_0)/\dot{V}(\mathbf{r},t_0)| \sim 650$ years.

Integrating both sides of Eq.(2) on a time interval $[t - \tfrac{1}{2} \Delta T \, ; \, t + \tfrac{1}{2} \Delta T)]$, with $\Delta T > T_{moon}$, gives

$$\int_{t-\tfrac{1}{2}\Delta T}^{t+\tfrac{1}{2}\Delta T} \Box\phi(\mathbf{r},t') \, dt' = (2/\mu_0\mathcal{F}) \int_{t-\tfrac{1}{2}\Delta T}^{t+\tfrac{1}{2}\Delta T} \mathbf{B}(\mathbf{r},t')^2 \, dt', \quad (9)$$

Now, let us consider the disturbances to the geomagnetic potential due to the Moon. There are two high tides and two low tides per day, strongly modulated on the lunar monthly spring/neap cycle. Now, the $T_{moon}$-periodic components vary slowly as compared to the $T_{tide}$-periodic components since the tidal period $T_{tide} \approx T_{moon}/55 \ll T_{moon}$, period of the orbital motion of the Moon around the Earth. Consequently, as the left-hand side of Eq.(9) implies derivatives, it is dominated by the rapidly varying term of the $T_{tide}$-periodic components. Conversely, the right-hand side of Eq.(9) does not imply derivatives, hence it is rather dominated by the slowly varying term of the $T_{moon}$-periodic components. Therefore, it follows in the first order approximation

$$\int_{t-\tfrac{1}{2}\Delta T}^{t+\tfrac{1}{2}\Delta T} \Box\phi(\mathbf{r},t') \, dt' = \Box\int_{t-\tfrac{1}{2}\Delta T}^{t+\tfrac{1}{2}\Delta T} \phi(\mathbf{r},t') \, dt' \approx \Box\int_{t-\tfrac{1}{2}T_{tide}}^{t+\tfrac{1}{2}T_{tide}} \phi(\mathbf{r},t') \, dt' \quad (10)$$

and

$$\int_{t-\tfrac{1}{2}\Delta T}^{t+\tfrac{1}{2}\Delta T} \mathbf{B}(\mathbf{r},t')^2 \, dt' \approx \int_{t-\tfrac{1}{2}T_{moon}}^{t+\tfrac{1}{2}T_{moon}} \mathbf{B}(\mathbf{r},t')^2 \, dt'. \quad (11)$$

Thus, Eq.(9) yields

$$T_{tide} \, \Box\langle\phi\rangle(\mathbf{r},t) \approx (2/\mu_0\mathcal{F}) \, \langle\mathbf{B}^2\rangle(\mathbf{r},t) \, T_{moon}, \quad (12)$$

or otherwise stated

$$\Box\langle\phi\rangle(\mathbf{r},t) \approx (T_{moon}/T_{tide}) \, (2/\mu_0\mathcal{F}) \, \langle\mathbf{B}^2\rangle(\mathbf{r},t), \quad (13)$$

with the average values $\langle\phi\rangle(\mathbf{r},t)$ and $\langle\mathbf{B}^2\rangle(\mathbf{r},t)$ defined as follows

$$\langle\phi\rangle(\mathbf{r},t) = \int_{t-\tfrac{1}{2}T_{tide}}^{t+\tfrac{1}{2}T_{tide}} \phi(\mathbf{r},t') \, dt'/T_{tide} \quad (14)$$

and $\langle\mathbf{B}^2\rangle(\mathbf{r},t) = \int_{t-\tfrac{1}{2}T_{moon}}^{t+\tfrac{1}{2}T_{moon}} \mathbf{B}(\mathbf{r},t')^2 \, dt'/T_{moon}. \quad (15)$

Now, in as much as $|\delta\phi| = |\phi - 1| \ll 1$, the average value of the effective gravitational constant at the epoch t, reads

$$\langle G_{eff}(\mathbf{r},t)\rangle = \langle G/\phi\rangle(\mathbf{r},t) = \langle G/(1+\delta\phi)\rangle(\mathbf{r},t) \approx \langle G(1-\delta\phi)\rangle(\mathbf{r},t) = G(1-\langle\delta\phi\rangle)(\mathbf{r},t)$$

$$\approx G/(1+\langle\delta\phi\rangle)(\mathbf{r},t) = (G/\langle 1+\delta\phi\rangle)(\mathbf{r},t) = G/\langle\phi\rangle(\mathbf{r},t). \quad (16)$$

Thus,

$$1/\langle G_{eff}(\mathbf{r},t)\rangle - 1/\langle G_{eff}(\mathbf{r},t_0)\rangle \approx -(T_{moon}/T_{tide})(2/G\mu_0\mathcal{F})V(\mathbf{r},t_0)\dot{V}(\mathbf{r},t_0)\Delta t. \quad (17)$$

III-Comparison with the experimental data

In China, the team led by J. Luo at the center for gravitational experiments of the HUST has been conducting continuously precision measurements of the gravitational constant since 1997. Recently, together with other colleagues from China and Russia, the HUST group has published two precise but discordant values of the gravitational constant compared with their previous measurements [30, 31]. One of the experiments uses the time-of-swing (TOS) technique, in which the pendulum oscillates. The frequency of oscillation is determined by the positions of the external masses and $G_{lab}$ can be deduced by comparing frequencies for two different mass configurations. The second experiment uses the angular-acceleration feedback (AAF) method, which involves rotating the external masses and the pendulum on two separate turntables. A feedback mechanism monitors the twist angle of the pendulum, which is held at zero by changing the angular speed of one of the turntables; $G_{lab}$ is then derived from the rate of change required to produce a zero angle. Since the HUST team carried out both methods on different apparatus and two different laboratories, one does not expect any correlation between the systematic errors involved in both methods. The distance between both laboratories is about 150 meters, so their coordinates are almost the same. By using Google maps [32], the HUST coordinates are : Latitude = 30.519, Longitude = 114.414 and Latitude = 1456 m (r = 6372.656 km, see Earth Radius by Latitude[5] WGS 84). The four values of $G_{lab}$ published so far by the HUST team and the epoch when they were respectively carried out are the following :

HUST 05 (see, [30]) : epoch $t_0$ = 1997, $G_{lab}$ = (6.672 3 ± 0.000 9) × $10^{-11}$ $m^3$ $kg^{-1}$ $s^{-2}$ ;

HUST 09 (see, [31]) : epoch t between 2007 and 2008 or between[6] 2006 and 2008, $G_{lab}$ = (6.673 49 ± 0.000 18) × $10^{-11}$ $m^3$ $kg^{-1}$ $s^{-2}$,

HUST 18 (see, [14, 32]) : epoch t between 2014 and 2017 (TOS method) and t between 2014 and 2018 (AAF method), $G_{lab}$ = (6.674184 ± 0.000078) × $10^{-11}$ $m^3$ $kg^{-1}$ $s^{-2}$ (TOS method) and $G_{lab}$ = (6.674484 ± 0.000078) × $10^{-11}$ $m^3$ $kg^{-1}$ $s^{-2}$ (AAF method).

---

[5] https://es.planetcalc.com/7721/
[6] https://www.nist.gov/sites/default/files/documents/pml/div684/fcdc/Jun_Luo-pdf.pdf

Hereafter, $\Delta t$ is expressed in years and we set $G_0 = 10^{-11}$ m$^{-3}$ kg$^1$ s$^2$.

Besides, $1/<G_{eff}(\mathbf{r},t_0)> = 1/G_{eff}(\mathbf{r},t_0) = 0.149873357\times10^{11}$ SI and $(2G_0/G\mu_0\mathcal{F})$ $V(\mathbf{r},t_0)$ $\dot{V}(\mathbf{r},t_0)$ = $2.41031\times10^{-6}$ m$^{-3}$ kg$^1$ s$^2$/yr (see table 1, FIG. 1 and FIG. 2 for the best fit of the slope). Therefore,

$$\mathcal{F}^{-1} \times (T_{moon}/T_{tide}) = [G\mu_0/G_0\times 2V(\mathbf{r},t_0)\dot{V}(\mathbf{r},t_0)]\times 2.41031\times 10^{-6}. \quad (18)$$

Now, $V(\mathbf{r},t_0) = 65.6680204297$ T.m and $-0.1187709643$ nT/yr $\leq \dot{V}(\mathbf{r},t_0) \leq -0.0955422055$ T.m/yr (see Table 1), so given that the sidereal orbital motion of the moon $T_{moon} \approx 27.321662$ days and the period of the tides $T_{tide} \approx 12.420556$ hours, one finds

$$\mathcal{F}^{-1} = (3.25 \pm 0.35)\times 10^{-14} \text{ m/J}. \quad (19)$$

As one can see, the above estimate matches the earlier one [1] recalled in section I.

| year | $\Delta t = t - t_0$ (year) | $g^0{}_1(t)$ (Gauss) | $g^1{}_1(t)$ (Gauss) | $h^1{}_1(t)$ (Gauss) | $V(\mathbf{r},t)$ (T.m) |
|---|---|---|---|---|---|
| 1995 | $-2$ | $-0.29692$ | $-0.01784$ | 0.05306 | $-65.4717403101$ |
| 2000 | 3 | $-0.296194$ | $-0.017282$ | 0.051861 | $-65.9624406093$ |
| 2005 | 8 | $-0.2955463$ | $-0.0166905$ | 0.0507799 | $-66.4271623646$ |
| 2010 | 13 | $-0.2949657$ | $-0.0158642$ | 0.0494426 | $-67.0948098063$ |
| 2015 | 18 | $-0.29442$ | $-0.01501$ | 0.047971 | $-67.8471595964$ |

Table 1 : The above data imply respectively for the time interval [1995 ; 2000], [1995 ; 2005] and [1995 ; 2015] : $<(dV/dt)(\mathbf{r},t)> = [V(\mathbf{r},2000) - V(\mathbf{r},1995)]/5 = -0.0981400598$ T.m/yr, $<(dV/dt)(\mathbf{r},t)> = [V(\mathbf{r},2005) - V(\mathbf{r},1995)]/10 = -0.0955422055$ T.m/yr and $<(dV/dt)(\mathbf{r},t)> = [V(\mathbf{r},2015) - V(\mathbf{r},1995)]/20 = -0.1187709643$ nT/yr.

The tables and figures below show the plots $G_0/G_{lab}(r_{HUST},t) - G_0/G_{lab}(r_{HUST},t_0)$ versus $\Delta t$ obtained by varying the epoch when measurements were carried out within the time intervals provided by the authors. All plots are consistent with a reference date $t_0$ being a given date in 1997 ± 0.30037 year (1997 ± 3 months 20 days).

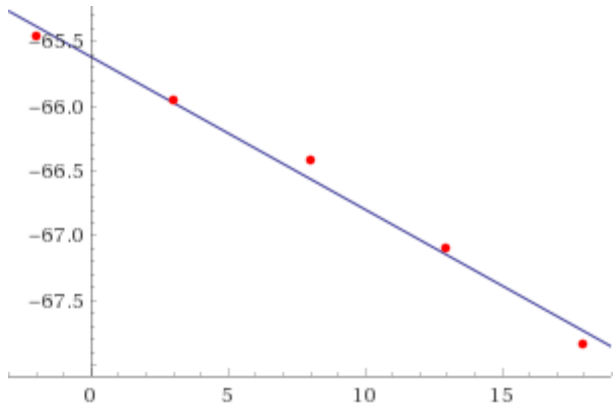

FIG. 1 : The linear least-squares best fit yields V(r,t) = − 0.117664155392 $\Delta t$ − 65.6193492942, with a correlation coefficient R = − 0.99418354, so that <(dV/dt)(r,t)> = − 0.117664155392 nT/m/yr.

Table 2

| $\Delta t$ (year) | 0 | 11 | 18 | 20 |
|---|---|---|---|---|
| $G_0/G_{lab}(r_{HUST},t) - G_0/G_{lab}(r_{HUST},t_0)$ | 0 | − 0.000026725 | − 0.0000423065 | − 0.000049041 |

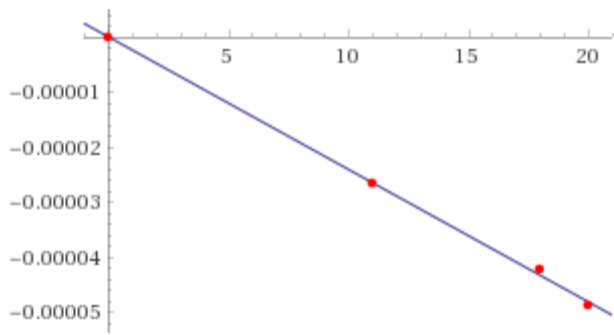

FIG. 2 : The least-squares fit to the data yields : $G_0/G_{lab}(r_{HUST},t) - G_0/G_{lab}(r_{HUST},t_0)$ = − 2.41031×10$^{-6}$ $\Delta t$, with a correlation coefficient R = − 0.999330 ; $t_0$ = 1997 + 0.003401077 yr.

Table 3

| Δt (year) | 0 | 11 | 17 | 20 |
|---|---|---|---|---|
| $G_0/G_{lab}(r_{HUST},t) - G_0/G_{lab}(r_{HUST},t_0)$ | 0 | − 0.000026725 | − 0.0000423065 | − 0.000049041 |

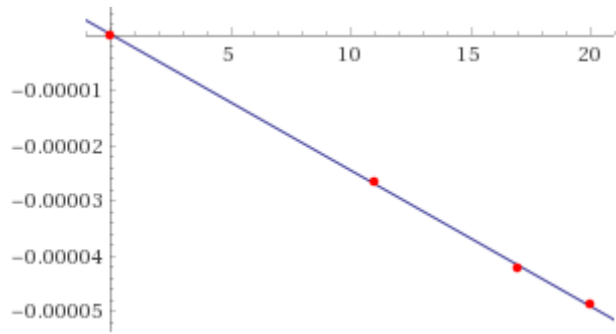

FIG. 3 : The least-squares fit to the data yields : $G_0/G_{lab}(r_{HUST},t) - G_0/G_{lab}(r_{HUST},t_0) = - 2.46639 \times 10^{-6} \Delta t$, with a correlation coefficient R = − 0.999156 ; $t_0$ = 1997 + 0.0318551811 yr.

Table 4

| Δt (year) | 0 | 11 | 17 | 19 |
|---|---|---|---|---|
| $G_0/G_{lab}(r_{HUST},t) - G_0/G_{lab}(r_{HUST},t_0)$ | 0 | − 0.000026725 | − 0.0000423065 | − 0.000049041 |

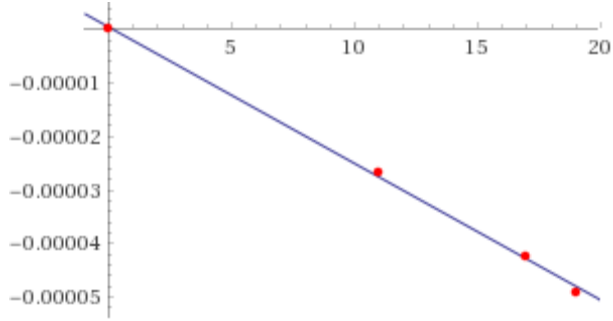

FIG. 4 : The least-squares fit to the data yields : $G_0/G_{lab}(r_{HUST},t) - G_0/G_{lab}(r_{HUST},t_0) = -2.54909 \times 10^{-6} \Delta t$, with a correlation coefficient R = $-0.999156$ ; $t_0$ = 1997 + 0.1701152176 yr.

Table 5

| $\Delta t$ (year) | 0 | 11 | 19 | 21 |
|---|---|---|---|---|
| $G_0/G_{lab}(r_{HUST},t) - G_0/G_{lab}(r_{HUST},t_0)$ | 0 | $-0.000026725$ | $-0.0000423065$ | $-0.000049041$ |

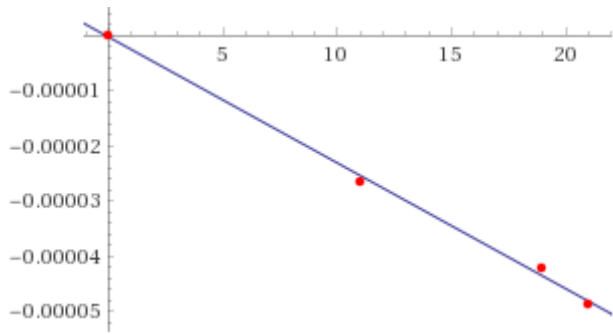

FIG. 5 : The least-squares fit to the data yields : $G_0/G_{lab}(r_{HUST},t) - G_0/G_{lab}(r_{HUST},t_0) = -2.28134 \times 10^{-6} \Delta t$, with a correlation coefficient R = $-0.998497$ ; $t_0$ = 1997 + 0.1889490387 yr.

Table 6

| Δt (year) | 0 | 11 | 16 | 18 |
|---|---|---|---|---|
| $G_0/G_{lab}(\mathbf{r}_{HUST},t) - G_0/G_{lab}(\mathbf{r}_{HUST},t_0)$ | 0 | − 0.000026725 | − 0.0000423065 | − 0.000049041 |

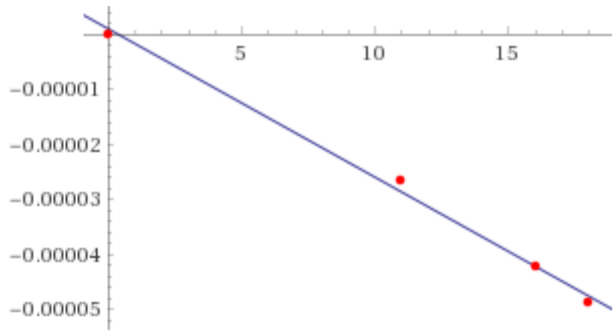

FIG. 6 : The least-squares fit to the data yields : $G_0/G_{lab}(\mathbf{r}_{HUST},t) - G_0/G_{lab}(\mathbf{r}_{HUST},t_0) = -2.69731\times10^{-6}\,\Delta t$, with a correlation coefficient R = − 0.99757 ; $t_0$ = 1997 + 0.3064634766 yr.

Table 7

| Δt (year) | 0 | 11 | 17 | 18 |
|---|---|---|---|---|
| $G_0/G_{lab}(\mathbf{r}_{HUST},t) - G_0/G_{lab}(\mathbf{r}_{HUST},t_0)$ | 0 | − 0.000026725 | − 0.0000423065 | − 0.000049041 |

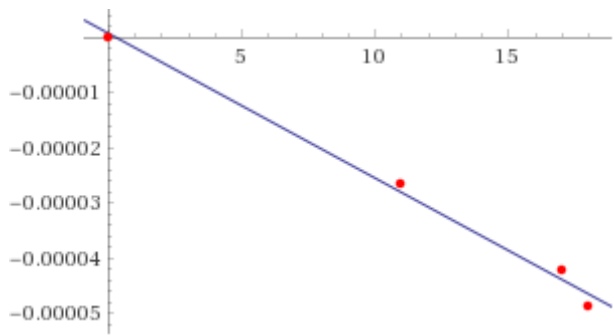

FIG. 7 : The least-squares fit to the data yields : $G_0/G_{lab}(\mathbf{r}_{HUST},t) - G_0/G_{lab}(\mathbf{r}_{HUST},t_0) = -2.62483\times10^{-6}\,\Delta t$, with a correlation coefficient R = − 0.995984 ; $t_0$ = 1997 + 0.2542633237 yr.

Table 8

| Δt (year) | 0 | 11 | 20 | 21 |
|---|---|---|---|---|
| $G_0/G_{lab}(r_{HUST},t) - G_0/G_{lab}(r_{HUST},t_0)$ | 0 | − 0.000026725 | − 0.0000423065 | − 0.000049041 |

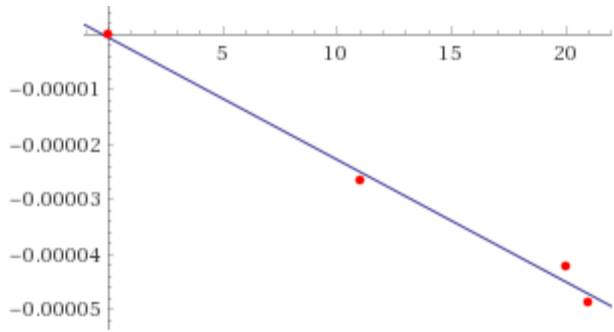

FIG. 8 : The least-squares fit to the data yields : $G_0/G_{lab}(r_{HUST},t) - G_0/G_{lab}(r_{HUST},t_0) = -2.22036 \times 10^{-6} \, \Delta t$, with a correlation coefficient R = − 0.99513 ; $t_0$ = 1997 − 0.2942856113 yr.

Table 9

| Δt (year) | 0 | 11 | 17 | 17 |
|---|---|---|---|---|
| $G_0/G_{lab}(r_{HUST},t) - G_0/G_{lab}(r_{HUST},t_0)$ | 0 | − 0.000026725 | − 0.0000423065 | − 0.000049041 |

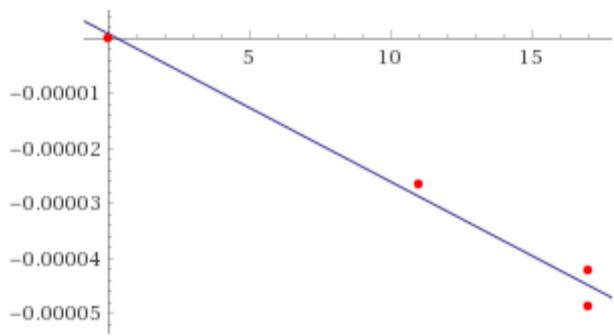

FIG. 9 : The least-squares fit to the data yields : $G_0/G_{lab}(r_{HUST},t) - G_0/G_{lab}(r_{HUST},t_0) = -2.69036 \times 10^{-6} \, \Delta t$, with a correlation coefficient R = − 0.98988 ; $t_0$ = 1997 + 0.2781873058 yr.

IV-Conclusion

We have addressed the ongoing question raised by the discordant values of the gravitational constant, $G_{lab}$, measured with great precision in the same laboratory but over different epochs spanning a large period of time. Focusing on the HUST measurements which cover the 1997-2018 period yielding four values differing from one another by more than 100 ppm, we have shown that a temporal variation introduces greater consistency into this apparent lack of concordance. According to the KKψ theory, this temporal variation is induced by the magnetic field of the Earth as source of the internal 5D KK scalar field which in turn masks the true gravitational constant by yielding an effective one. Thus, the measurement of the true gravitational constant, G, can be achieved on space-based laboratories by getting rid of the geomagnetic field and the magnetic fields of the other planets and the sun. In these conditions, a precision of 10 ppm can be reached beyond an altitude of 20 089 km from the surface of the Earth that is above the orbits of the current global navigation satellite systems in operation, namely the GPS and GLONASS.

V-Appendix

$\Delta V(\mathbf{r},t_0) = 0$, so that $\Delta \dot{V}(\mathbf{r},t_0) = 0$, since $\Delta V(\mathbf{r},t) = 0$ at any time t. In addition,

$\phi(\mathbf{r},t) \approx \phi(\mathbf{r},t_0) + \phi^{(1)}(\mathbf{r},t_0) \Delta t + \frac{1}{2} \phi^{(2)}(\mathbf{r},t_0) \Delta t^2$,

at the second order in $\Delta t$. Further,

$\partial^2 \phi(\mathbf{r},t)/\partial t^2 \approx \phi^{(2)}(\mathbf{r},t_0)$,

$\Box \phi(\mathbf{r},t) = \phi^{(2)}(\mathbf{r},t_0)/c^2 - \Delta \phi(\mathbf{r},t_0) - \Delta \phi^{(1)}(\mathbf{r},t_0) \Delta t - \frac{1}{2} \Delta \phi^{(2)}(\mathbf{r},t_0) \Delta t^2$,

$2K(\nabla V(\mathbf{r},t))^2/\mu_0 = 2\mathcal{F}^{-1} (\nabla V(\mathbf{r},t_0))^2/\mu_0 + 4\mathcal{F}^{-1} \nabla V(\mathbf{r},t_0) \cdot \nabla \dot{V}(\mathbf{r},t_0) \Delta t/\mu_0 + 2\mathcal{F}^{-1} (\nabla \dot{V}(\mathbf{r},t_0))^2 \Delta t^2/\mu_0$.

Therefore,

$\phi^{(2)}(\mathbf{r},t_0)/c^2 - \Delta \phi(\mathbf{r},t_0) - \Delta \phi^{(1)}(\mathbf{r},t_0) \Delta t - \frac{1}{2} \Delta \phi^{(2)}(\mathbf{r},t_0) \Delta t^2 = 2\mathcal{F}^{-1} (\nabla V(\mathbf{r},t_0))^2/\mu_0$

$+ 4\mathcal{F}^{-1} \nabla V(\mathbf{r},t_0) \cdot \nabla \dot{V}(\mathbf{r},t_0) \Delta t/\mu_0 + 2\mathcal{F}^{-1} (\nabla F(\mathbf{r},t_0))^2 \Delta t^2/\mu_0$,

so that

$\Delta \phi(\mathbf{r},t_0) = \phi^{(2)}(\mathbf{r},t_0)/c^2 - 2\mathcal{F}^{-1} (\nabla V(\mathbf{r},t_0))^2/\mu_0$,

$\Delta \phi^{(1)}(\mathbf{r},t_0) = - 4\mathcal{F}^{-1} \nabla V(\mathbf{r},t_0) \cdot \nabla \dot{V}(\mathbf{r},t_0)/\mu_0$,

$$\Delta\phi^{(2)}(\mathbf{r},t_0) = -4\mathcal{F}^{-1}(\nabla V(\mathbf{r},t_0))^2/\mu_0,$$

the solutions of which read

$$\phi^{(1)}(\mathbf{r},t_0) = -2\mathcal{F}^{-1}\dot{V}(\mathbf{r},t_0)V(\mathbf{r},t_0)/\mu_0 \text{ and } \phi^{(2)}(\mathbf{r},t_0) = -2\mathcal{F}^{-1}V(\mathbf{r},t_0)^2/\mu_0,$$

hence,

$$\Delta\phi(\mathbf{r},t_0) = -(2/\mu_0\mathcal{F})[\dot{V}(\mathbf{r},t_0)^2/c^2 + (\nabla V(\mathbf{r},t_0))^2] \approx -2\mathcal{F}^{-1}(\nabla V(\mathbf{r},t_0))^2/\mu_0,$$

and then

$$\phi(\mathbf{r},t_0) \approx 1 - \mathcal{F}^{-1}V(\mathbf{r},t_0)^2/\mu_0,$$

$$\phi(\mathbf{r},t) \approx 1 - \mathcal{F}^{-1}V(\mathbf{r},t_0)^2/\mu_0 - 2\mathcal{F}^{-1}V(\mathbf{r},t_0)\dot{V}(\mathbf{r},t_0)\Delta t/\mu_0 - \mathcal{F}^{-1}\dot{V}(\mathbf{r},t_0)^2\Delta t^2/\mu_0$$

$$= 1 - \mathcal{F}^{-1}[V(\mathbf{r},t_0)^2 + 2V(\mathbf{r},t_0)\dot{V}(\mathbf{r},t_0)\Delta t + \dot{V}(\mathbf{r},t_0)^2\Delta t^2]/\mu_0 + b\,\Delta t + a\,\Delta t^2$$

$$= 1 - \mathcal{F}^{-1}V(\mathbf{r},t)^2/\mu_0 + b\,\Delta t + a\,\Delta t^2.$$

Since $\lim_{r\to\infty}\phi(\mathbf{r},t) = 1$, it follows $b = a = 0$, one obtains $\phi(\mathbf{r},t) \approx 1 - \mathcal{F}^{-1}V(\mathbf{r},t)^2/\mu_0$.